\documentclass{aa} 
\usepackage{graphicx} 
\usepackage{times,float}

\begin{document}

\title{\object{GRB 030227}: the first multiwavelength afterglow of an 
INTEGRAL GRB\thanks
{Based on observations with INTEGRAL, an ESA project 
with instruments and science data centre 
funded by ESA member states (especially the PI
countries: Denmark, France, Germany, Italy, Switzerland, Spain), Czech
Republic and Poland, and with the participation of Russia and the USA.
Also partially based on observations collected by the Gamma-Ray Burst 
Collaboration at ESO (GRACE) at the European Southern Observatory, Chile 
(ESO Large Programme 165.H-0464).} 
}

\author{A. J.    Castro-Tirado      \inst{1} 
   \and J.       Gorosabel          \inst{1,2} 
   \and S.       Guziy              \inst{1} 
   \and D.       Reverte            \inst{1} 
   \and J. M.    Castro Cer\'on     \inst{2} 
   \and A.       de Ugarte Postigo  \inst{1} 
   \and N.       Tanvir             \inst{3} 
   \and S.       Mereghetti         \inst{4}  
   \and A.       Tiengo             \inst{4}  
   \and J.       Buckle             \inst{5}
   \and R.       Sagar              \inst{6} 
   \and S. B.    Pandey             \inst{6}
   \and V.       Mohan              \inst{6}    
   \and N.       Masetti            \inst{7}       
   \and F.       Mannucci           \inst{8}
   \and S.       Feltzing           \inst{9}
   \and I.       Lundstrom          \inst{9} 
   \and H.       Pedersen           \inst{10} 
   \and C.       Riess              \inst{11}
   \and S.       Trushkin           \inst{12} 
   \and J.       V\'{\i}lchez       \inst{1}
   \and N.       Lund               \inst{13} 
   \and S.       Brandt             \inst{13} 
   \and S.       Mart\'{\i}nez N\'u\~nez \inst{14} 
   \and V.       Reglero            \inst{14}  
   \and M. D.    P\'erez-Ram\'\i rez \inst{15}
   \and S.       Klose              \inst{16}    
   \and J.       Greiner            \inst{17} 
   \and J.       Hjorth             \inst{10} 
   \and L.       Kaper              \inst{18} 
   \and E.       Pian               \inst{19} 
   \and E.       Palazzi            \inst{7}    
   \and M. I.    Andersen           \inst{20} 
   \and A.       Fruchter           \inst{2} 
   \and J. P. U. Fynbo              \inst{21}
   \and B. L.    Jensen             \inst{10}  
   \and C.       Kouveliotou        \inst{22} 
   \and J.       Rhoads             \inst{2}
   \and E.       Rol                \inst{18} 
   \and P. M.    Vreeswijk          \inst{23} 
   \and R. A. M. J. Wijers             \inst{17} 
   \and E.       van den Heuvel     \inst{17} 
   }

\offprints{A.J. Castro-Tirado, \email{ajct@iaa.es} }

\institute{Instituto de Astrof\'\i sica de Andaluc\'\i a (IAA-CSIC), Apartado de Correos, 3004, 18080 Granada, Spain. 
       \and Space Telescope Science Institute, 3700 San Mart\'\i n Drive, Baltimore MD 21218, USA. 
       \and Department of Physical Sciences, University of Hertfordshire, College Lane, Hatfield, Herts AL10 9AB, UK. 
      \and Istituto di Astrofisica Spaziale e Fisica Cosmica, Sez. di Milano ``G. Occhialini'' - CNR v. Bassini 15, I-20133 Milano, Italy. 
      \and Joint Astronomy Centre, 660 N, A\' \rm ohoku Place, Hilo, HI 96720, 
           USA.
      \and State Observatory, Manora Peak, Naini Tal 263129, Uttaranchal, India.
      \and Istituto di Astrofisica Spaziale e Fisica Cosmica -- Sezione di 
           Bologna, Via Gobetti 101, 40129 Bologna, Italy. 
      \and Istituto di Radioastronomia, sezione di Firenze, CNR, Largo E. Fermi 5, 50125, Firenze, Italy.
      \and Lund Observatory, S\"olvegatan 27, P.O. Box 43, SE-221 00 Lund, 
           Sweden.
      \and Astronomical Observatory, University of Copenhagen, Juliane Maries Vej 30, 2100 Copenhagen \O\ , Denmark. 
      \and   University Observatory Munich, Scheinerstr. 1, 81679 Munich, Germany.
      \and Special Astrophysical Observatory of R.A.S., Karachai-Cherkessia, 
369167, Nizhnij Arkhyz, Russia. 
      \and Danish Space Research Institute, Juliane Maries Vej 30, 2100 Copenhagen \O\ , Denmark. 
      \and Grupo de Astrof\'{\i}sica y Ciencias del Espacio, Universidad de Valencia, Apdo. 2085, 46071 Burjassot (Valencia), Spain. 
      \and European Space Agency, Research Science Division, Noorwijk, The Netherlands.
      \and Th\"uringer Landessternwarte Tautenburg, 07778 Tautenburg, Germany. 
      \and Max Planck Institut f\"ur extraterrestrische Physik, Giessenbachstr., PF 1312, 85741 Garching, Germany. 
      \and University of Amsterdam, Kruislaan 403, 1098 SJ Amsterdam, The Netherlands. 
      \and Osservatorio Astronomico di Trieste, Via Tiepolo 11, 34131 Trieste, Italy. 
      \and Astrophysikalisches Institut, An der Sternwarte 16, 14482 Potsdam, Germany. 
      \and Department of Physics and Astronomy, Aarhus University, Ny  
           Munkegade, 8000 Aarhus C, Denmark. 
      \and NASA/MSFC, NSSTC, SD-50, 320 Sparkman Drive, Huntsville, AL 35805, USA. 
      \and European Southern Observatory, Alonso de C\'ordova 3107, Casilla 19001, Santiago 19, Chile. 
           }

\date{Received / Accepted } 
 
\abstract{We present multiwavelength observations of a  
  gamma-ray burst detected by INTEGRAL (\object{GRB 030227}) between 5.3   
  hours and $\sim$  1.7 days after the  
  event. Here we report the discovery of a dim optical afterglow (OA) that 
  would not have been detected by many previous searches due to its 
  faintess (R$\sim$ 23). This OA was seen to decline following a 
  power  law  decay with index   $\alpha_\mathrm{R}$ = $-$0.95 $\pm$ 0.16.  
  The spectral  index $\beta_\mathrm{opt/NIR}$  yielded $-$1.25 $\pm$ 0.14. 
  These values may be  
  explained by a relativistic expansion of a fireball (with $p$ = 2.0) in the 
  cooling regime. 
  We also find evidence for inverse Compton scattering in X-rays.

         } 
 
\maketitle 
 
\section{Introduction} 
 
Gamma  Ray Bursts (GRBs) are  flashes of  high energy  ($\sim$ 1 
keV--10 GeV)  photons (\cite{fishman-meegan95}), occurring  at cosmological 
distances. Since their discovery in 1967, $\sim$ 
3\,000 GRBs  have been detected  in $\gamma$ rays but only $\sim$ 50 have 
been pinpointed at optical wavelengths in the last six years (Greiner 2003), 
with redshifts  ranging  from $z$ = 
0.0085 (\cite{galama98}) to $z$  = 4.50 (\cite{andersen00}). 
 
ESA's INTEGRAL satellite offers unique capabilities for the detection  
of GRBs thanks to its high sensitivity and imaging capabilities at 
$\gamma$-ray energies and X-ray/optical. \object{GRB 030227} was discovered 
in the INTEGRAL IBIS data
by the automatic IBAS software (Mereghetti et al. 2003a) on 27 February 
2003 (see Fig. 1).
The burst started at 08:42:03 UT and lasted for $\approx18$~s, putting it 
in the ``long-duration'' class of GRBs. 
It had  a peak flux of 1.1 photons cm$^{-2}$ s$^{-1}$ and a fluence of 
7.5$\times10^{-7}$   erg cm$^{-2}$ in the 20-200 keV range (Mereghetti et al. 
2003b). The prompt dissemination (50 min) of the GRB position
(G\"{o}tz et al. 2003a,b) enabled triggering of a target of opportunity 
(ToO) observation with ESA's XMM-Newton satellite, starting $\sim$8 hr 
after the event. This observation revealed a fading X-ray source consistent 
with the IBIS error circle which was identified as the X-ray afterglow 
of GRB 030227 (Loiseau et al. 2003, Gonz\'alez-Riestra et al. 2003).  
The INTEGRAL data analysis indicates a soft gamma-ray spectrum for GRB  
030227, possibly placing it in the  ``X-ray rich'' class, with the brightest 
X-ray afterglow detected by XMM-Newton so far (Mereghetti et al. 2003b).  
Here we report the discovery of the optical afterglow of the GRB, 
and present further multiwavelength observations.

\begin{figure} 
\begin{center}
      \resizebox{7.6cm}{!}{\includegraphics{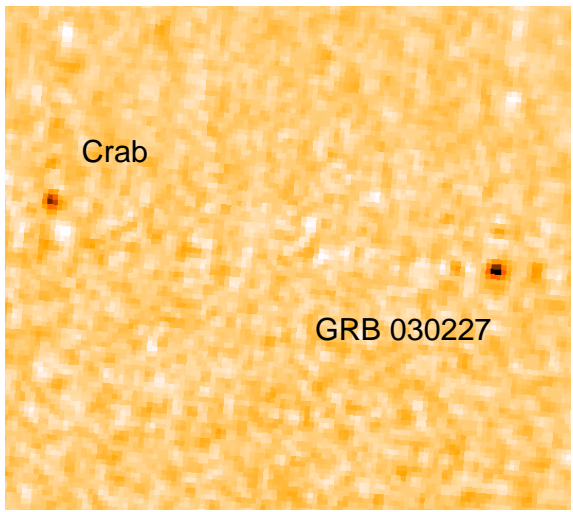}} 
        \caption{INTEGRAL image of GRB030227 in the Crab field during 
                 the calibration phase. The image has been obtained in 
                 the 15-200 keV range with the IBIS/ISGRI instrument and 
                 deconvolved using the IBAS software. The distance between 
                 the Crab and the GRB is $\sim$ 10$^{\circ}$. North is up,
                 East to the left.}
    \label{secuencia temporal}
\end{center} 
\end{figure}

\section{Observations} 
  \label{observaciones}

  ToO  observations   were   triggered   starting    
  5.3 hr after the event at the 1.0 m State Observatory  
  telescope (1.0SO) in Nainital (India). Subsequently, observations were 
  made at the Wendelstein 0.8 m telescope (0.8Wend) close to Munich (Germany), 
  at the 2.5 m Isaac Newton  
  Telescope (2.5INT), at the 2.56 m Nordic Optical Telescope (2.56NOT) 
  and the 3.58 m Telescopio Nazionale  Galileo (3.58TNG), at 
  La Palma (Spain),  and at the 3.6 m telescope (3.6ESO) at the  
  European  Southern Observatory, at La Silla (Chile).  
  Near-IR observations were obtained at the 3.8m United Kingdom Infrared  
  Telescope (UKIRT) on Mauna Kea (Hawaii).
  Table \ref{tabla1} displays the observing log. 
  We    performed  aperture photometry     using  the  PHOT    routine under 
  IRAF\footnote{IRAF  is   distributed   by the   NRAO,  
  which are operated by USRA, 
  under  cooperative  agreement with  the US 
  NSF.}.  The optical field was  calibrated  using  
  the secondary standard stars provided by Henden (2003). The near-IR images 
  were calibrated using the standard FS114.
  The RATAN-600 radio telescope in  
  Karachai-Cherkessia (Russia), observed the field at 3.9~GHz  
  during   the period Mar 6 -- 9, 2003.

\begin{table*} 
      \begin{center} 
            \caption{Journal of optical and near-infrared (NIR) observations of the \object{GRB 030227} field.} 
                     \begin{tabular}{@{}lccccc@{}} 
 
Date of 2003 UT          & Telescope  & Filter & Exposure Time & Magnitude  \\ 
                         &            &        &    (seconds)  &            \\ 
 
\hline

Feb 27, 13:45--14:00 & 1.0SO (CCD)      & $R$ &   300 + 600    & 22.0 $\pm$ 0.3 \\ 
Feb 27, 18:08--18:40 & 0.8Wend (MONICA) & $R$ &   1\,320       & $>$21.5    \\ 
Feb 27, 20:40--20:50 & 2.5INT (WFC)     & $R$ &    600         & 23.10 $\pm$ 0.16 \\ 
Feb 27, 20:51--21:01 & 2.5INT (WFC)     & $B$ &    600         & 24.5 $\pm$ 0.3 \\ 
Feb 27, 23:17--23:30 & 2.5INT (WFC)     & $R$ &    800         & 23.19 $\pm$ 0.18 \\ 
Feb 27, 23:32--23:45 & 2.5INT (WFC)     & $B$ &    800         &  $>$23.0 \\ 
Feb 28, 05:00--05:40 & 3.8UKIRT (UFTI) & $K$ &   1\,800       & 19.2 $\pm$ 0.1 \\ 
Feb 28, 05:40--06:30 & 3.8UKIRT (UFTI) & $H$ &   1\,800       & 20.2 $\pm$ 0.1 \\ 
Feb 28, 20:00--20:47 & 2.56NOT (ALFOSC) & $R$ & 3 $\times$ 600 & 24.1 $\pm$ 0.2 \\ 
Feb 28, 20:10--21:00 & 3.5TNG (DOLORES) & $B$ & 3 $\times$ 900 & 25.4 $\pm$ 0.2 \\ 
Feb 28, 22:25--23:15 & 2.5INT (WFC)     & $R$ & 3 $\times$ 600 & $>$ 23.9    \\ 
Feb 28, 23:15--Mar 1, 00:37 & 3.6ESO (EFOSC2) & $R$ & 6 $\times$ 600 & 24.75 $\pm$ 0.25 \\ 
\hline 
                         \label{tabla1} 
                     \end{tabular} 
      \end{center} 
\end{table*}

\begin{figure} 
\begin{center}
      \resizebox{7.0cm}{!}{\includegraphics{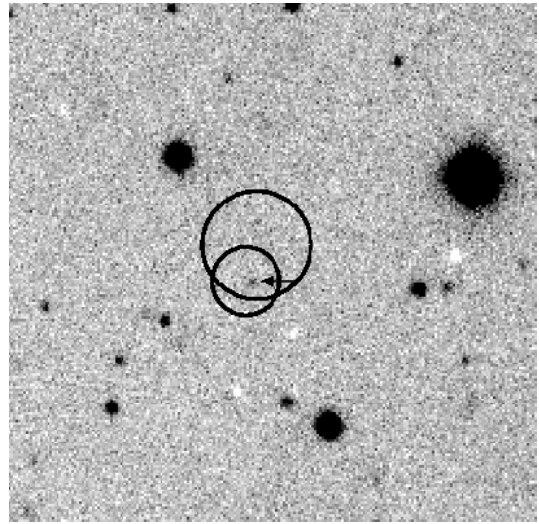}} 
      \caption{The discovery $R$ band image of the \object{GRB 030227} field  
               taken at the 2.5INT on 27 Feb 2003. The position of the  
               OA is indicated by the arrow inside the preliminary (Loiseau 
               et al. 2003) and final (Mereghetti et al. 2003b) XMM-Newton 
               error circles (6$^{\prime\prime}$ and 4$^{\prime\prime}$ radii, 
               respectively). The field is $1^{\prime} \times 1^{\prime}$ with 
               North up and East to the left.} 
    \label{carta ID}
\end{center} 
\end{figure}

\section{Results and discussion} 
  \label{resultados} 
   
  The deep optical observations taken at the 2.5INT (simultaneously with 
  the XMM-Newton follow-up) revealed a point source within the 
  6$^{\prime\prime}$ radius XMM-Newton error circle, which was proposed as 
  the optical afterglow (OA) to GRB 030227 (Castro-Tirado et al. 2003;  
  see Fig. 2).  
  This identification was confirmed by further optical imaging 
  (Soderberg et al. 2003, Berger et al. 2003, Gorosabel et al. 2003), which 
  showed the source to be fading. An  astrometric solution based   
  on 10 USNO A2-0 reference stars in the 2.5INT image taken on  
  27 Mar 2003   yields    for   the    OA    $\alpha_\mathrm{2000}$   = 
  4$^\mathrm{h}$57$^\mathrm{m}$33.05$^\mathrm{s}$,  $\delta_\mathrm{2000}$ = 
  $+$20\degr29\arcmin05.0$\arcsec$.  The internal  error of the position is 
  0.55\arcsec, which has  to be added to the  $1\sigma$ systematic error of 
  the    USNO    catalogue    ($\simeq$    0.25$\arcsec$    according    to 
  Assafin et al. 2001). The final astrometric error corresponds to 
  0.60$\arcsec$.   
  No radio afterglow was detected with RATAN-600 at the position of the 
  OA down to a flux density limit (3$\sigma$) of 2 mJy (at 3.9 GHz).

    \subsection{The lightcurve of the \object{GRB 030227} OA} 
             \label{curva de luz} 
              
             Our $R$  band lightcurve (Fig. 3) 
             shows that the source was  declining in brightness.            
             Most GRB optical  counterparts appear to be well characterised 
             by a  power  law decay plus a   constant flux component,  $F(t) 
             \propto  (t  -  t_\mathrm{0})^\alpha  +F_\mathrm{host}$ where, 
             $F(t)$ is  the total measured  flux of the counterpart at time 
             ($t  -  t_\mathrm{0}$)  after   the  onset  of  the  event  at 
             $t_\mathrm{0}$,  $\alpha$   is the  temporal decay   index and 
             $F_\mathrm{host}$ is  the flux of the  underlying host galaxy.  
             $F_\mathrm{host}$  is negligible in this case,   
             as indicated by  
             the lack of flattening of the optical light curve at the  
             time of our last observations (we estimate $R_{\rm host}>26$). 
             From a least    squares  linear regression   to   the 
             observed $R$ band fluxes, a power  law  decline  with   
             $\alpha_\mathrm{R}$  = $-$1.10   $\pm$  0.14 
             ($\chi^2$/dof = 1.53) provides an acceptable fit  
             but it is ruled out by the early upper limits. 
             However, if we exclude the 3.6ESO data point, a power law  
             decline with $\alpha_\mathrm{R}$  = $-$0.95   $\pm$  0.16 
             ($\chi^2$/dof = 0.64) provides a better fit, which is  
             also just consistent with the earlier upper limits. 
             This  flux decay  of  the \object{GRB  030227}  agrees with the 
             decay in X-rays  $\alpha_\mathrm{X}$  $\simeq$ $-$1.0 $\pm$ 0.1 
             (Mereghetti et al. 2003b, Watson et al. 2003) and is 
             comparable to the slow   
             decline rates observed in other GRB OAs.   
             There are OAs for which a break or a smooth, gradual transition  
             does occur within 1-2 days of the burst. 
             The 3.6ESO measurement might give indication of a break around  
             $\sim$ 1.5 days but the lack of further data at later epochs 
             precludes  confirmation. 
 
     An upper limit to the redshift of $z$ $\leq$ 3.5 can be estimated from 
     the absence of the onset of the Lyman forest blanketing in the optical 
     data, consistent with the estimates from the X-ray spectra  
     ($z$ $\sim$ 3-4, Mereghetti et al. 2003b, $z$ $\sim$ 1.6, Watson et al. 
     2003).

 \begin{figure}
 \begin{center} 
 \resizebox{8.0cm}{5.0cm}{\includegraphics{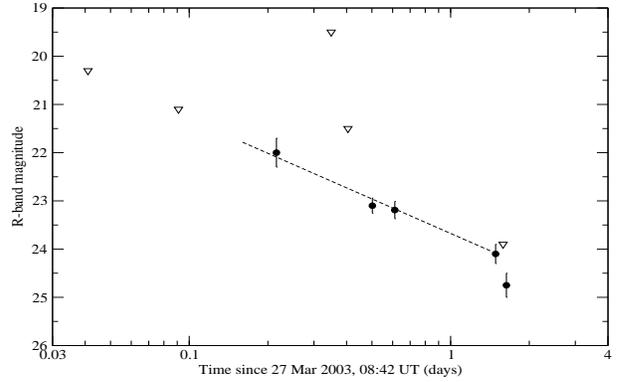}} 
 
 \caption{     
   The $R$  band lightcurve of the \object{GRB 
   030227} OA.  Circles represent measured magnitudes and triangles represent  
   upper   limits.   The dashed line is the best fit to the data (excluding 
   the 3.6ESO point), a power-law  
   decline with  $\alpha_\mathrm{R}$ = $-$0.95 $\pm$ 0.16. In addition to 
   upper limits reported in this paper we also include 
   the earlier ones derived by Urata et al.  
   (2003), Pavlenko et al. (2003) and Torii et al. (2003).} 
 
    \label{curvas de luz}
  \end{center} 
\end{figure}

 \subsection{The spectral shape of the afterglow: evidence for inverse Compton 
      scattering} 
      \label{forma espectral} 
       
      We have  determined the spectral flux distribution of the 
      \object{GRB 030227} 
      OA on 28.24 UT Feb 2003 (mean epoch of the $HK$ band images) 
      by  means of our $BRHK$ broad  band  photometric measurements  
      obtained with the different telescopes.  We interpolated the B \& R 
      band magnitudes to that epoch, and fitted the  observed  flux  
      distribution with a power  law $F_\mathrm{\nu} \propto  \nu^\beta$, 
      where $F_\mathrm{\nu}$ is the flux density at frequency $\nu$, 
      and $\beta$ is the spectral index. The optical flux densities at the 
      wavelengths of $BRHK$  bands have been   
      derived without subtracting the  contribution  of any host  galaxy,  
      assuming a reddening $E(B - V)$ = 
      0.46 from the  DIRBE/IRAS  dust   maps   (\cite{schlegel98}).  
      In converting 
      the magnitude into flux, the effective wavelengths and normalisations 
      given in Fukugita et al. (1995) were used. The flux densities are 
      2.4, 2.7, 10.2 and 15.0 $\mu$Jy at the $BRHK$  
      bands, corrected for 
       Galactic reddening (but not  for possible intrinsic absorption in 
      the host  galaxy).   The fit  to  the NIR/optical data  $F_\mathrm{\nu}  
      \propto \nu^\beta$ gives $\beta_\mathrm{opt/NIR}$ = $-$1.25 $\pm$ 0.14 
      ($\chi^{2}$/dof = 0.08).

\begin{figure*} 
\begin{center}
  \resizebox{12cm}{6cm}{\includegraphics{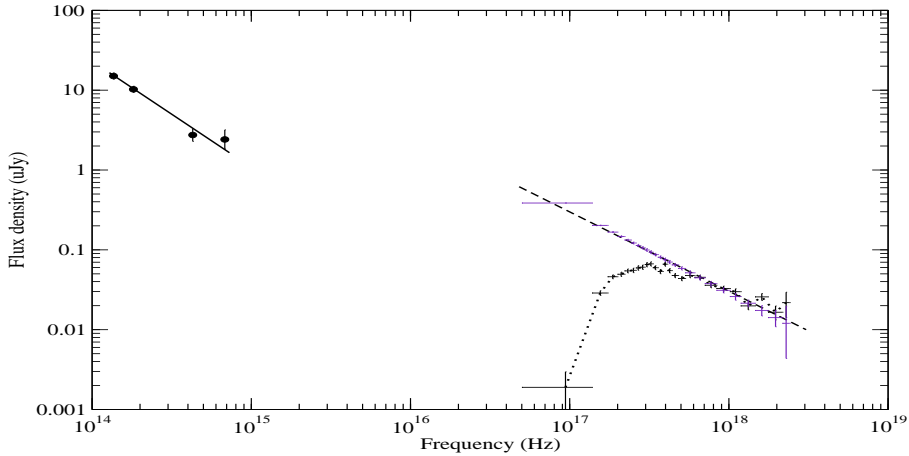}} 
 
\caption{The broadband spectrum   
         for the \object{GRB 030227} afterglow at $t_\mathrm{0}$ + 0.87 days. 
         The NIR/optical spectrum (solid line) has been dereddened  
         assuming $E(B-V)$ = 0.46 from the DIRBE/IRAS dust maps. A fit   
         $F_\mathrm{\nu}  \propto \nu^\beta$ to the NIR/optical data  
         gives $\beta_\mathrm{opt/NIR}$ = $-$1.25 $\pm$ 0.14.
         The absorbed X-ray spectrum from XMM-Newton (dotted line) can  
         be unabsorbed and represented by a power law (dashed line) with 
         photon index $\Gamma$ = 1.94 $\pm$ 0.05  
         (i.e. $\beta_\mathrm{X}$ = $-$0.94 $\pm$ 0.05)
         , considering $N_\mathrm{H}$ =  
         6.8 $\times$ 10$^{22}$ cm$^{-2}$ at a redshift z $\sim$ 4, 
         following Mereghetti et al. (2003b). This is equivalent to consider 
         a value of $N_\mathrm{H}$ =  1.1 $\times$ 10$^{22}$ cm$^{-2}$ at $z$ 
         $\sim$ 1.6, as derived by and Watson et al.(2003).} 
    \label{espectro}
\end{center} 
\end{figure*}

      Fig. 4 shows the broadband spectrum (from near-IR/optical photometry to  
      X-rays) for the \object{GRB 030227} afterglow. The NIR/optical spectral 
      index ($\beta_\mathrm{opt/NIR}$ = $-$1.25 $\pm$ 0.14) is consistent with 
      the spectral index for the unabsorbed X-ray spectrum 
      ($\beta_\mathrm{X}$ = $-$0.94$\pm$0.05, Mereghetti et al. 2003b) 
      but they do not 
      match each other\' \rm s extrapolations, similarly to GRB 000926 
      (Harrison et al. 2001) and GRB 010222 (in\' \rm t Zand  et al. 2001). 
      We have investigated whether considerable extinction  in the host 
      galaxy (considering different extinction  laws) could produce such 
      effect, in order to reproduce  an optical-IR spectrum as an 
      extrapolation of the X-ray  spectrum, but found this not feasible.
      This results in no spectral break between the NIR/optical and X-ray 
      bands, which will be also consistent with the similar decay indexes in 
      both bands ($\alpha_\mathrm{R}$ = $-$0.95 and $\alpha_\mathrm{X}$ = 
      $-$1.0). Thus, we suggest that in contrast to the NIR/optical band, where
      synchrotron processes dominate, there is an important contribution 
      of inverse Compton scattering to the X-ray spectrum besides line 
      emission (Watson et al. 2003), as it has been 
      proposed for GRB 000926 (Harrison et al. 2001). 
      This implies a lower limit on the density of the external medium, 
      $n$ $\geq$ 10 cm$^{-3}$ (see Panaitescu and Kumar 2000).

        \subsection{Adiabatic expansion or cooling regime ?} 
        \label{models}

        Many afterglows exhibit a single power law decay index. Generally this 
        index  is  $\alpha$  $\sim$  $-$1.3,  a reasonable  value  for  the 
        spherical  expansion of  a relativistic  blast wave  in  a constant 
        density  interstellar medium, according to the standard fireball 
        model  (\cite{meszaros-rees97}).  
        In fact, the  value of  $\alpha$ for GRB 030227 falls  within   the   
        boundaries  defined  by   
        the observations made  to date, from $-$0.67  $\pm$ 0.1 in  the  
        \object{GRB 020331}  (Dullighan et al. 2002)  to  
        $-$1.73  $\pm$ 0.04 in the \object{GRB 980519} (\cite{jaunsen01}).
        $\beta$ only depends on $p$ (the exponent of the power-law 
        distribution of the Lorentz factor for the relativistic electrons) 
        and it does not depend on the geometry of the expansion.  
        Hereafter we will assume $\alpha$ $\simeq$ $-$1.0  and 
        $\beta$ $\simeq$ $-$1.0 for the NIR/optical/X-ray bands.
        Several models have been explored in order to reproduce the observed  
        values of $\alpha$ and $\beta$. 

        For an adiabatic expansion  ($\nu_\mathrm{m} < \nu  < \nu_\mathrm{c}$) 
        in a constant 
        density insterstellar medium (ISM), $\beta$ = (1 $- p$)/2  and 
        $\alpha$ = $-$3($p -$1)/4, 
        where $\nu$ is the observing frequency, $\nu_\mathrm{c}$ is the 
        cooling break frequency and $\nu_\mathrm{m}$ is the synchrotron 
        peak frequency (\cite{sari98}).  
        For a spherical  adiabatic  expansion  with the density $n$ $\propto$ 
        r$^{-s}$ with $s$ =  2 (inhomogeneous medium due to a stellar wind, 
        Chevalier and Li 2000), $\beta$ = (1 $- p$)/2  and 
        $\alpha$ = $-$(3$p -$1)/4.
        In both cases, we have considered values of $p$ in the range 1.8 to 
        3, as observed in most afterglows detected to date 
        (\cite{van-paradijs00}), but we cannot reproduce the observed values 
        of $\alpha$ and $\beta$.

        For the cooling regime ($\nu_\mathrm{c} < \nu$) in both the ISM (s = 0)
        and  wind (s = 2) cases, the evolution is similar at 
        NIR/optical and X-ray wavelengths, with $\beta$ = $- p$/2  and 
        $\alpha$ = $-$(3$p -$2)/4.  The best results are obtained for  
        $p$  $\simeq$ 2.0, from which we derive $\alpha$  $\simeq$ $-$1.0 
        and $\beta$ $\simeq$ $-$1.0, consistent with our 
        measurements.  
         
        In light of the previous arguments, we propose that both the observed 
        slow decay in the NIR/opt/X-ray lightcurves and the intrinsic 
        spectrum, are consistent with a fireball in the cooling regime with 
        $p$ = 2.0, but we cannot distinguish between the s = 0 and s = 2 
        cases. Only a detection in radio ($\nu < \nu_\mathrm{c}$ at 
        $t_\mathrm{0}$ + 0.87 days) would have allowed us to discriminate 
        both models.

\section{Conclusions} 
  \label{conclusiones} 
   
  We presented  multiwavelength observations of the afterglow associated with 
  the possibly X-ray rich \object{GRB 030227}.
  This would be one of the few OAs detected to date to an X-ray rich GRB.
  The decay index in the $R$-band lightcurve is 
  $\alpha_\mathrm{R}$ = $-$0.95 $\pm$ 0.16, with a possible break detection 
  at $t_\mathrm{0}$ + $\sim$ 1.5 days.    
  The optical-NIR spectrum at $t_\mathrm{0}$ + 0.87 days allowed us  
  to determine a spectral index $\beta_\mathrm{opt/NIR}$ = $-$1.25 $\pm$ 0.14.
  The multiwavelength spectrum can be modeled by the expansion of
  a fireball (with $p$ = 2.0) in the cooling regime.  
  We also found evidence for inverse Compton scattering in X-rays.
 
  The GRB 030227 OA is only 0.5 mag brighter that the dimmest OAs found 
  so far, like GRB980613 (Hjorth et al. 2002), GRB 000630 (Fynbo et al. 2001), 
  GRB 020322 (Bloom et al. 2002) 
  and GRB 021211 (Fox et al. 2003), once the galactic extinction 
  is taken into account. 
  For comparison purposes  see the light curves  
  of 18 afterglows shown in Fig. 3 of Gorosabel et al. (2002). 

  In combination with multiwavelength studies, it is expected that INTEGRAL 
  will shed more light on the origin of GRBs.

\begin{acknowledgements} 
   
  We thank the Comit\'e de Asignaci\'on de Tiempos  
  del Observatorio del Roque de los Muchachos 
  at Canary Islands (Spain) for generous allocation of the Spanish  
  ToO programme.
  The data presented  here have been taken in part 
  using  ALFOSC, which  is owned  by the IAA and operated at the NOT under 
  agreement between IAA and the  NBIfAFG of the Astronomical Observatory of 
  Copenhagen. Part of the observations presented  in this paper were 
  obtained  
  under the ESO Programme 70.D-0227 (granted to the GRACE team). 
  This work is also partially based on data taken with
  the UKIRT operated by the JAC on behalf PPARC. 
  This research has been partially supported by the Ministerio de
  Ciencia y Tecnolog\'{\i}a de Espa\~na under the programme AYA2002-0802  
  (including FEDER funds). 
 
\end{acknowledgements}

\end{document}